\documentclass[a4paper,11pt]{article}
\usepackage{pos}
\usepackage{comment}

\title{Obstacles for dS in Supersymmetric Theories}
\author*[a]{Maxim Emelin}
\affiliation[a]{Department of Physics "Galileo Galilei", University of Padua,\\
  via Marzolo 8, Padua, Italy}

\emailAdd{maxim.emelin@pd.infn.it}
\abstract{We present two types of obstacles for constructing de Sitter space in supersymmetric theories that appear as effective theories of quantum gravity. The first obstacle concerns de Sitter critical points in extended supergravity with charged massless gravitini. We show that all such critical points have a Hubble scale of the same order or above the UV cutoff dictated by the magnetic Weak Gravity Conjecture. The second obstacle concerns the dynamics of the goldstino in models with non-linearly realized supersymmetry, such as string models involving anti-branes. We show, using an exact renormalization group approach, that the goldstino sector has an instability toward the formation of condensates, which may have important qualitative consequences for de Sitter constructions involving anti-brane uplifts. }

\FullConference{%
  Corfu Summer Institute 2021 "School and Workshops on Elementary Particle Physics and Gravity"\\
  29 August - 9 October 2021\\
  Corfu, Greece
}

\begin{document}
\maketitle

\section{Introduction}

The task of constructing (quasi-)de Sitter configurations in string theory is an important topic of study both for its cosmological applications as well as for the general understanding of the structure of string theory and quantum gravity. The feasibility of de Sitter vacua in string theory remains a point of contention, with various proposals and objections, but no complete and explicit constructions to date \cite{Danielsson:2018ztv}.

At least part of the challenge in constructing de Sitter space in string theory can be attributed to the fact that String theory is intrinsically supersymmetric, while de Sitter space is a positive energy configuration and therefore intrinsically non-supersymmetric. This means that any construction of de Sitter space within string theory must involve some form of supersymmetry breaking and various questions of stability and control of corrections must be approached with great care. This is, of course, also true of other non-supersymmetric scenarios within string theory.

One possible approach to non-supersymmetric scenarios in string theory is to work directly with a system where supersymmetry is broken already at the string scale \cite{Antoniadis:1999xk,Mourad:2017rrl}, however this gives up much of the computational control that supersymmetry offers and much remains to be understood in such systems. A more common approach is to have the supersymmetry breaking occur at lower energies where a lower-dimensional effective theory description is available. The form of the effective theory is then restricted by supersymmetry and the supersymmetry breaking itself can be understood in terms of specific string theory ingredients of the compactification. 

The possible sources of supersymmetry breaking in 4-d supergravity theories can come in the form of gauging global symmetries, superpotentials or explicit matter sectors with non-linear supersymmetry realizations. Of these options, the first is available in minimal as well as extended supergravity, and can be realized through internal fluxes in type II Calabi-Yau string compactifications \cite{Polchinski:1995sm,Michelson:1996pn,Taylor:1999ii}. Superpotentials are of course only present in $\mathcal{N}=1$ supersymmetric theories and also trace their origins to internal fluxes \cite{Gukov:1999ya} as well as possible non-perturbative effects \cite{Witten:1996bn}. Non-linear realizations of supersymmetry, in the form of constrained superfields \cite{Lindstrom:1979kq,Kapustnikov:1981de,Samuel:1982uh,Bergshoeff:2015tra,Cribiori:2017ngp,DallAgata:2016syy,Ferrara:2014kva}, occur on the worldvolumes of anti-branes \cite{Bergshoeff:2015jxa,Bandos:2015xnf,Dasgupta:2016prs,Cribiori:2019bfx,Cribiori:2020bgt} and are a staple ingredient in most proposed stable de Sitter vacuum constructions \cite{Kachru:2003aw,Kachru:2003sx,Balasubramanian:2005zx,Bena:2022cwb}.

In this contribution, we will explore some obstacles to constructing controlled de Sitter states using these ingredients. $\mathcal{N}=1$ superpotentials are the least constrained by any general physical principles, so we will not discuss them here. In section \ref{sec2}, we will present constraints that arise from combining the strict relationship between gauging in $\mathcal{N}=2$ supergravity to the scalar potential with the magnetic Weak Gravity Conjecture, originally studied in \cite{Cribiori:2020use,DallAgata:2021nnr}. The results reveal an incompatibility between the Hubble scale of de Sitter critical points with charged but massless gravitini and the UV cutoff dictated by the conjecture. In section \ref{sec3} we will turn our attention to the dynamics of the non-linear supersymmetry sector described by constrained superfields, and summarize the approach and results of \cite{DallAgata:2022abm}, where evidence of a previously unnoticed instability toward the formation of goldstino condensates was found. This instability potentially greatly alters the expected properties of string theory configurations involving anti-branes and, depending on its eventual endpoint, calls for a re-examination of many existing de Sitter constructions.

\section{Extended Supersymmetry and the Weak Gravity Conjecture} \label{sec2}

The scalar potential in supergravity theories with extended supersymmetry arises entirely from the gauging. The values of the energy at critical points of the scalar potential is therefore related to the values of the gauge couplings. On the other hand, general quantum gravity considerations, stemming primarily from the dynamics of charged black holes, produce some of the most robust constraints on effective theories that admit a quantum gravity UV completion. One of the most established such constraint is the Weak Gravity Conjecture (WGC), which can be obtained from the requirement that extremal black holes should be able to decay \cite{Arkani-Hamed:2006emk}. This implies the existence of particles with $U(1)$ charge greater than their mass (the electric Weak Gravity Conjecture), but also implies that the cutoff of any consistent effective theory with an unbroken $U(1)$ gauge symmetry with coupling $g$ is bounded by
\begin{equation}
 \Lambda_{UV} \leq g q M_p 
\end{equation}
for every non-vanishing charge $q$. The latter constraint is called the magnetic Weak Gravity Conjecture (mWGC). In \cite{Cribiori:2020use} it was further argued that the mWGC can be used to constrain de Sitter critical points and quasi-de Sitter configurations, by further demanding that the Hubble scale lies parametrically below the mWGC cutoff.
\begin{equation}
    H \ll \Lambda_{UV}
\end{equation}
This criterion can be motivated by the observation that, since powers of the Hubble scale govern the typical magnitude of higher curvature terms in the action, $H/\Lambda_{UV}$ appears as the EFT expansion parameter. Furthermore, the presence of thermal fluctuations in de Sitter space, whose size is similarly given by $H$, demanding that these fluctuations do not push the system outside the EFT regime of validity similarly requires parametric separation between the Hubble scale and the mWGC cutoff.

The above argument, combined with the connection between the scalar potential and gauging in extended supergravity, opens the door for some strict constraints on de Sitter critical points in these theories. Indeed in \cite{Cribiori:2020use} it was noticed that many of the de Sitter critical points that one can obtain in models containing only vector multiplets have a Hubble scale on the order of the mWGC cutoff and therefore suffer a breakdown of their EFT description. It was further noted that these examples involved a vanishing gravitino mass matrix, and so it was conjectured that massless gravitini could serve as a signal for the breakdown of effective field theory descriptions. This idea was further elaborated in \cite{Cribiori:2021gbf,Castellano:2021yye} connecting massless gravitini with the swampland distance conjecture \cite{Ooguri:2006in}. For a more detailed discussion of the connections between de Sitter space, the mWGC and other swampland conjectures, we refer to the contribution to these proceedings by Niccol\`o Cribiori \cite{Cribiori:2022sxf} and references therein. Finally, in \cite{DallAgata:2021nnr}, the analysis of \cite{Cribiori:2020use} was extended to models with both vector and hypermultiplets and a rigorous proof was given that any de Sitter critical point in $\mathcal{N}=2$ supergravity with charged massless gravitini will have an energy on the order of or above the cutoff dictated by the mWGC.

In the rest of this section we will review the general proof of the main result as well as give two examples one of which illustrates the result in action while the other provides an example of de Sitter critical points that evade exclusion on mWGC grounds, due to the lack of an unbroken $U(1)$ and a non-vanishing gravitino mass matrix. Both examples also illustrate some subtleties related to possible gauge group enhancements at certain points in field space.

\subsection{The general result}

Here we prove that de Sitter critical point in $\mathcal{N}=2$ supergravity with charged massless gravitini do not have parametric separation between the Hubble scale and the mWGC cutoff. Recall that the gravitino mass and charge matrices can be written as
\begin{equation} \label{gravMQ}
    \begin{aligned}
    S_{ij} &= i P^x_\Lambda L^\Lambda (\sigma_x)_i^k \epsilon_{jk}  \\
(Q_A)_i^{\ j} &= \frac{1}{2} \mathcal{E}^\Lambda_A \big( P^0_\Lambda \delta_i^{\ j} + P^x_\Lambda (\sigma^x)_i^{\ j} \big)
    \end{aligned}
\end{equation}
where $L^\Lambda$ denotes the choice of covariantly holomorphic section that specifies the scalar manifold for the vector multiplets and $P_\Lambda^{0,x}$ denote the prepotentials, which are computed from the killing vectors that specify the scalar manifold isometries gauged by the available vector fields. The ``gauge vielbein" $\mathcal{E}^\Lambda_A$ defines the canonically normalized charges and is related to the gauge-kinetic matrix $\mathcal{I}$ by
\begin{equation}
    \mathcal{E}_A^\Lambda \mathcal{E}_B^\Sigma \delta^{AB} = -\mathcal{I}^{-1|\Lambda \Sigma}
\end{equation}
This means that the eigenvalues of $Q_A$ are the physical gravitino charges, which are to be used when applying the mWGC.
Meanwhile, the scalar potential is a sum of three terms
\begin{equation}
\mathcal{V} = \mathcal{V}_1 + \mathcal{V}_2 + \mathcal{V}_3 
\end{equation}
with
\begin{equation} \label{potterms}
\begin{aligned}
\mathcal{V}_1 &= g_{I\bar{J}} k^I_\Lambda k^{\bar{J}}_\Sigma \bar{L}^\Lambda L^\Sigma \\ %= g^{I\bar{J}} U_I^M \bar{U}_{\bar{J}}^N P_M^0 P_N^0 \ , \qquad U_I^\Lambda = D_I L^\Lambda \nonumber \\
\mathcal{V}_2 &= 4 h_{uv} k^u_\Lambda k^v_\Sigma L^\Lambda \bar{L}^\Sigma  \\
\mathcal{V}_3 &= (U^{\Lambda \Sigma} - 3 L^\Lambda \bar{L}^\Sigma ) P^x_\Lambda P^x_\Sigma
\end{aligned}
\end{equation}
where $g_{IJ}$ is the metric on the scalar manifold for the vector multiplets, $h_{uv}$ is the metric on the scalar manifold of the hypermultiplets and $U^{\Lambda \Sigma} = g^{I\bar{J}} \nabla_I L^\Lambda \nabla_{\bar{J}} \bar{L}^\Sigma = -\frac{1}{2} (\mathcal{I}^{-1})^{\Lambda \Sigma} - \bar{L}^\Lambda L^\Sigma$.
A vanishing mass matrix $S_ij$ requires $P^x_\Lambda L^\Lambda = 0$. This simplifies the expression for the scalar potential to
\begin{equation}
    \mathcal{V} = -\frac{1}{2}\mathcal{I}^{-1|\Lambda \Sigma} \big( P^0_\Lambda P^0_\Sigma + P^x_\Lambda P^x_\Sigma \big) + 4 h_{uv} k^u_\Lambda k^v_\Sigma L^\Lambda \bar{L}^\Sigma
\end{equation}
where in the first term we used the relations between the killing vectors $k_\Lambda^I$, their corresponding prepotentials $P_\Lambda^0$, and the gauge-kinetic matrix to rewrite $\mathcal{V}_1$. $\mathcal{V}_3$ itself simplifies to a similar looking form, but involving the prepotentials associated to the hypermultiplet scalar manifold. The last term is simply $\mathcal{V}_2$ and is positive definite. Since we are interested in finding a lower bound on the scalar potential we can drop this term and write
\begin{equation}
       \mathcal{V} \geq \frac{1}{2} \delta^{AB} \big( P^0_A P^0_B + P^x_A P^x_B \big)= \frac{1}{2} \delta^{AB} \big(\delta_i^j P_A^0 + \sigma^x \ _i^j P_A^x \big) \big(\delta_i^j P_B^0 + \sigma^x \ _i^j P_B^x \big)
  \end{equation}
where we traded the gauge indices $\Lambda, \Sigma$ for their ``flat" counterparts $A,B$ using $\mathcal{E}_\Lambda^A$ that we introduced earlier. In the second equality, we re-expressed the sum over the $0,x$ in terms of the traces of products of Pauli matrices (along with the identity). We recognize the terms in the parentheses as the expressions for the gravitino charge matrix given in \eqref{gravMQ}. In particular, we can choose $\mathcal{E}_\Lambda^A$ such that the $U(1)$ charge that we intend to use for the mWGC lies along the $A=1$ direction and we can thus write the lower bound
\begin{equation}
    \mathcal{V} \geq \frac{1}{2} \text{Tr} (Q_1 Q_1) = q_1^2 + q_2^2
\end{equation}
where $q_1$ and $q_2$ are the physical charges of the individual gravitini, which themselves provide an upper bound on the UV cutoff from the mWGC $q_1^2+q_2^2\geq \Lambda_{UV}^2$. The final conclusion is that we have
\begin{equation}
\mathcal{V} \geq q_1^2 + q_2^2 \geq \Lambda_{UV}^2 \implies H \geq \Lambda_{UV}/\sqrt{3} 
\end{equation}
Thus the Hubble scale associated to the value of the potential is bounded below by an energy of order the mWGC cutoff, invalidating the EFT description of such critical points. \footnote{An analogous proof for $\mathcal{N}=8$ supergravity was also given in \cite{DallAgata:2021nnr}. One can reasonably expect that the result holds for all extended supergravity theories, but the full proof is currently unavailable.
}

A couple of remarks are in order here. First, the proof assumes that the full mass matrix vanishes, i.e. that both gravitini are massless. This assumption completely removes the negative contributions to the potential, allowing us to place the lower bounds in the way that we do. Second, the proof does {\it not} assume that we are precisely at a critical point of the potential, except for the last step where the Hubble scale is determined from the potential. This means that the conclusion should also be valid for quasi-de Sitter backgrounds, as long as the notion of a cosmological horizon and a corresponding Hubble scale makes sense. We will see both these considerations at play in the second example presented in the next subsection. 

Finally an important note is that our result is contained within the ``Festina-Lente" bound, which places a lower bound on the masses of all charged particles in a de Sitter background. This bound can be derived from considering the evaporation process of large black holes in de Sitter space and its connection to the Weak Gravity Conjecture has also been discussed \cite{Montero:2019ekk,Montero:2020rpl,Montero:2021otb}. It is a non-trivial cross check that at least a sub-statement of this bound can also be derived purely on mWGC grounds.

\subsection{Examples and Caveats}

As an illustrative example of our result at work, as well as of its caveats and possible extensions we will present two very similar models, which have the same matter content, but differ in the gauging. A distinguishing feature of the first example is a flat direction that tunes the gravitino mass, interpolating between regions that respect or violate the mWGC. The second example has several critical points, which exhibit either non-abelian enhancement of the gauge group, or its complete breaking, both of which are obstacles to applying the mWGC. Additional examples with interesting properties can be found in \cite{Cribiori:2020use,DallAgata:2021nnr}, including fully stable de Sitter critical points, whose EFT validity are however ruled out by the mWGC. Earlier works on finding de Sitter critical points in extended supergravity include \cite{Fre:2002pd,DallAgata:2012plb,Catino:2013}

The models we are going to consider will have the following scalar manifolds
\begin{equation}
\mathcal{M}_{SK} = \frac{\text{SU(1,3)}}{\text{SU(3)}\times \text{U(1)}} \ , \qquad  \mathcal{M}_{QK} = \frac{\text{SO(4,2)}}{\text{SO(4)}\times \text{SO(2)}} \, ,
\end{equation}
with holomorphic section
\begin{equation}
    Z = \begin{pmatrix} 1 \\ z^I \\ -i/2 \\ i z^I / 2 \end{pmatrix} \ , \qquad I = 1,2,3 \ .
\end{equation}
There is an $SO(3)$ isometry of $\mathcal{M}_{SK}$, which rotates the $z^I$ into each other, given by the killing vectors
\begin{equation}
    \kappa_1^I = \begin{pmatrix} 0 \\ z_3 \\ - z_2  \end{pmatrix} \ , \quad  \kappa_2^I = \begin{pmatrix} -z_3 \\ 0 \\ z_1  \end{pmatrix} \ , \quad  \kappa_3^I = \begin{pmatrix} z_2 \\ -z_1 \\ 0  \end{pmatrix}
\end{equation}
while $\mathcal{M}_{QK}$ has an $SO(3)$ isometry generated by $T_{\underline{12}} , T_{\underline{13}} , T_{\underline{23}}$, an $O(1,1)$ generated by $T_{\underline{46}}$ and a $U(1)$ isometry generated by $T_{\underline{56}}$, where 
\begin{equation}
		(T_{\underline{ab}})_{\underline{c}}{}^{\underline{d}} = \eta_{\underline{c}[\underline{a}} \delta_{\underline{b}]}^{\underline{d}}
\end{equation}
are the $SO(4,2)$ generators used in the construction of $\mathcal{M}_{QK}$.

The $\mathcal{M}_{SK}$ isometries are gauged by the three vector multiplets as
\begin{equation}
k_\Lambda^I = e_1\, \big( 0 \ , \ \kappa_1^I \ , \ \kappa_2^I \ , \ \kappa_3^I \big) \,. 
\end{equation}
and the SO(3)$\times$O(1,1) isometry of $\mathcal{M}_{QK}$ are gauged as
\begin{equation}
\begin{aligned}
k_\Lambda^u = \big( e_0 k_{T_{\underline{46}}}^u \ , \ e_1 k_{T_{\underline{12}}}^u \ , \ e_1 k_{T_{\underline{13}}}^u \ , \ e_1 k_{T_{\underline{23}}}^u \big).
\end{aligned}
\end{equation}
i.e. the $SO(3)$ isometry is still gauged by the three vector multiplets, with the same charge $e_1$ as the $\mathcal{M}_{SK}$ isometries, while the $O(1,1)$ is gauged by the graviphoton with charge $e_0$.
The corresponding prepotentials given by
\begin{equation} \label{prepot}
\begin{aligned}
P_0^x = \begin{pmatrix} 0 \\ 0 \\ 0 \end{pmatrix} \ , \quad P_1^x = \begin{pmatrix} e_1 \\ 0 \\ 0 \end{pmatrix} \ , \quad P_2^x = \begin{pmatrix} 0 \\ e_1 \\ 0 \end{pmatrix} \ , \quad P_3^x = \begin{pmatrix} 0 \\ 0 \\ e_1 \end{pmatrix} \ ,
\end{aligned}
\end{equation}

\begin{figure}
\includegraphics[scale=0.6]{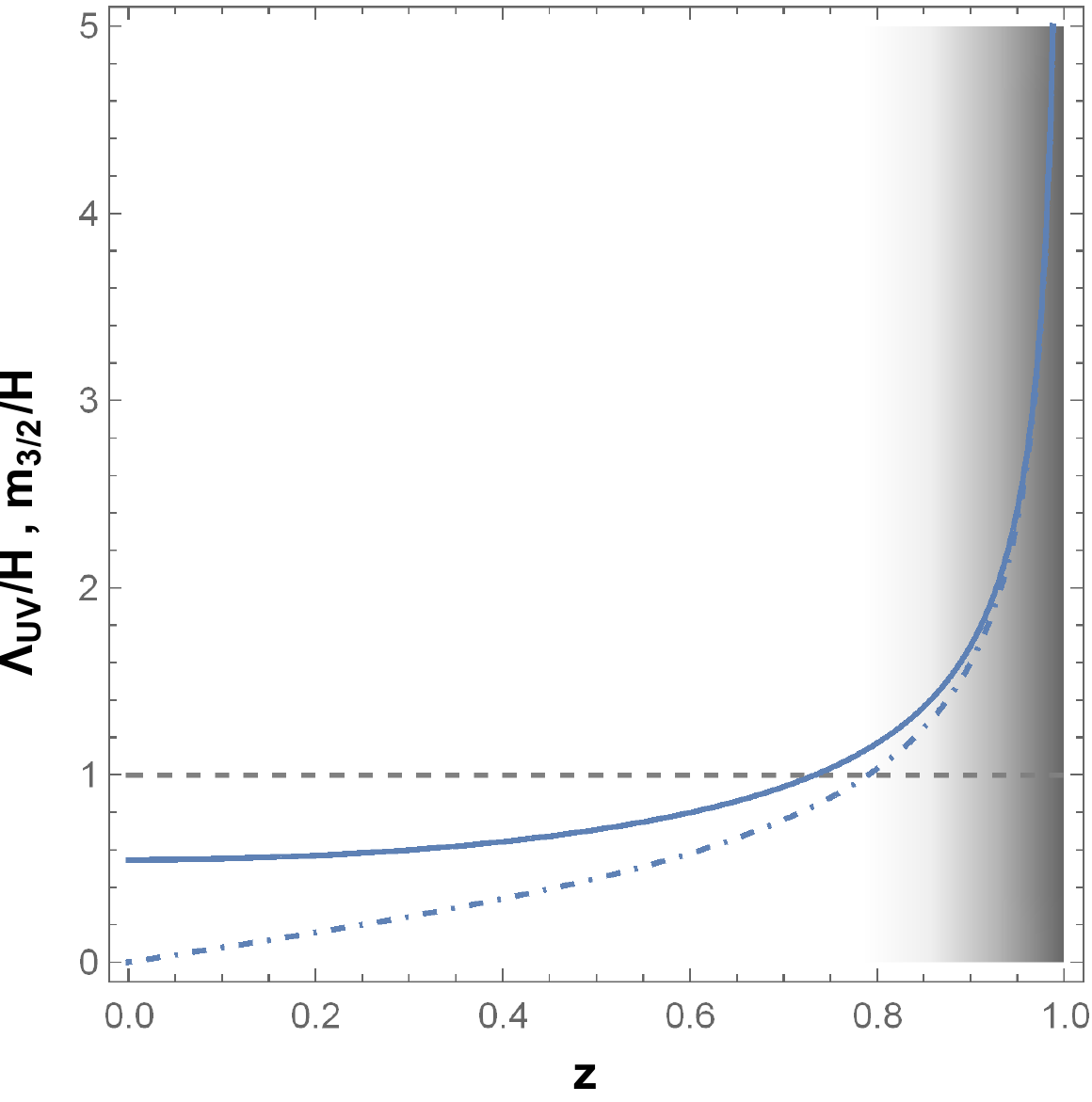}
\centering
\caption{The ratios $\Lambda_{UV}/H$ (solid) and $m_{3/2}/H$ (dot-dashed) as a function of the modulus $z$. For small values of $z$ the gravitino mass vanishes and the
Hubble scale is above the cut-off. The shaded gray
region denotes where the effective theory is increasingly well controlled; in the dark gray part
$H \ll \Lambda_{UV}$ . The gravitino mass is always below the cut-off, approaching it as $z$ approaches the
boundary of moduli space.} 
\label{fig1} 
\end{figure}

The resulting scalar potential has a critical point at the center of field space with
\begin{equation}
\begin{aligned}
{\cal V} = 2\, e_0^2 + 3\, e_1^2 \,
\end{aligned}
\end{equation}
and scalar masses related to the value of the potential as
\begin{align}
	\begin{split}
		m^2_{(mult.)} &= \left( 0_{(1)} \ , \ 2(e_0^2 - e_1^2)_{(3)} \ , \ 4 e^2_{0 \ (1)} \ , \ 4 e^2_{1 \ (2)} \right. \ , \\ &  \quad x_{1 \ {(1)}} \ , \ x_{2 \ {(1)}} \ , \ x_{3 \ {(1)}}  \\[2mm]
	& \quad e_1^2 + \sqrt{4 e_0^4 - 4 e_0^2 e_1^2 + 9 e_1^4}_{(2)} \ , \\ & \left.  \ e_1^2 - \sqrt{4 e_0^4 - 4 e_0^2 e_1^2 + 9 e_1^4}_{(2)} \right) \times {\cal V} \,
	\end{split}
\end{align}
where $x_1,x_2,x_3$ are the solutions to 
\begin{equation}
    x^3+6e_1 x^2 + (4 e_0^2 e_1^2-4e_0^4)x - (16e_0^4e_1^2 - 16e_0^2e_1^4+32e_1^6)=0
\end{equation}
Note that the critical point is unstable and its mass spectrum marginally respects the de Sitter conjecture.
If we further choose $e_0=e_1$ the scalar potential develops a flat direction along the imaginary directions of each of the $z^I$. Moving along any one of these flat directions will break the $SO(3)$ isometry to a $U(1)$, with the corresponding gravitino charge given by
\begin{equation} \label{gravCharge1}
    q_{U(1)}=\pm\sqrt{\frac{1+z^2}{1-z^2}}
\end{equation}
where $z$ is the magnitude of the imaginary part of the scalar that parametrizes our chosen flat direction. This $U(1)$ physical charge, and therefore the mWGC cutoff, interpolates between $1$ and infinity as $z$ grows from $0$ to $1$.
Meanwhile, the gravitino mass along this direction is given by
\begin{equation}
    S_{ij} = \begin{pmatrix} e_1\frac{z}{\sqrt{1-z^2}} & 0 \\ 0 & e_1 \frac{ z}{\sqrt{1-z^2}} \end{pmatrix}
\end{equation}
which vanishes as $z\to 0$ and diverges as $z \to 1$. This flat direction thus offers a perfect opportunity to illustrate the onset of the mWGC violation as we approach the massless point. Indeed, computing the mWGC cutoff in units of the Hubble scale as a function of $z$ we can see that the well-controlled regime appears precisely as $z\to 1$, where the gravitini are heavy, while the massless gravitino limit has the Hubble scale on the same order as the mWGC cutoff (see Figure \ref{fig1}).
Finally we also note that at $z=0$ the unbroken gauge symmetry gets enhanced to $SO(3)$, which is non-abelian and so the mWGC does not directly apply. However, since points along the flat directions are excluded by the mWGC arbitrarily close to the central critical point, we may conclude by continuity that the central point is similarly excluded.

As our second example, we may choose instead to gauge the $U(1)$ isometry on $\mathcal{M}_{QK}$ instead of the $O(1,1)$, i.e. choose
\begin{equation}
\begin{aligned}
k_\Lambda^u = \big( e_0 k_{T_{\underline{56}}}^u \ , \ e_1 k_{T_{\underline{12}}}^u \ , \ e_1 k_{T_{\underline{13}}}^u \ , \ e_1 k_{T_{\underline{23}}}^u \big).
\end{aligned}
\end{equation}
with the rest of the model exactly as before. In particular the prepotentials are still given by \eqref{prepot}. With this choice of gauging, the central critical point still exists and has energy
\begin{equation}
    \mathcal{V}=3 e_1^2
\end{equation}
with the mass spectrum 
\begin{equation}
    m^2_{(mult.)}=\left(-\frac{2}{3}_{(6)} \ , \ \frac{4}{3}r^2_{(2)} \ , \ \frac{4}{3}(r^2+1)_{(6)}\right) \times {\cal V}
\end{equation}
The vanishing of $P^x_0$ implies that the gravitini are uncharged under the separate $U(1)$ isometry. They do have charges $\pm \frac{1}{\sqrt{2}} e_1$ under the three generators of the $SO(3)$ isometry. However this isometry is non-abelian, so it isn't clear that the mWGC constraint can be applied. Furthermore, we no longer have the flat direction that allows us to break it to a single $U(1)$.

A possible way around this problem is to consider time-dependent trajectories, which have as initial boundary conditions points near, but not quite at, the central critical point, in such a way that the $SO(3)$ symmetry is broken down to $U(1)$ as in the previous example. In this case, the charge under this $U(1)$ will still be approximately equal to $\pm \frac{1}{\sqrt{2}} e_1$ and the energy, at least for some short time before the runaway due to the tachyonic directions takes over, will be above the mWGC cutoff, so we expect the EFT description around these trajectories to break down. By considering a limit of such runaway trajectories such that their initial conditions approach the critical point, we might wish to again conclude by continuity that the critical point itself also suffers a breakdown of its EFT description.

An important caveat to the above argument is that since the masses of the tachyons are comparable to the energy itself and therefore does not satisfy slow-roll conditions, it isn't clear that the mWGC criterion applies to such trajectories in the first place, since one can not meaningfully assign a Hubble radius to this solution. This question deserves further investigation.

Finally, a second interesting feature of this model is the presence of a second family of critical points located at 
\begin{equation}
    \text{Re}\, z_I = \pm\frac{1}{2} \ , \ \text{Im}\, z_J = \pm\frac{1}{2}  \, I\neq J \ .
\end{equation}
This critical point completely breaks the $SO(3)$ gauge symmetry, and the gravitino $U(1)$ charge continutes to vanish. The scalar mass spectrum is
\begin{equation}
\begin{aligned}
m^2_{(mult.)}&=\left(0_{(3)}, -1_{(2)} , 8_{(1)},2+4r^2-2r_{(2)},\beta^2+\beta_{(2)}, \beta^2 - \beta_{(2)} , 2+2r+4r^2_{(2)}\right) \times {\cal V} \, 
\end{aligned}
\end{equation}
and thus again has tachyons satisfying the de Sitter criterion. This critical point can therefore not be excluded by this criterion, nor by the mWGC due to a lack of any unbroken gauge group. The reason this is not a counterexample to our general theorem is that one of the gravitini becomes massive. Indeed the mass matrix evaluated at this critical point is
\begin{equation}
S_{ij} = \begin{pmatrix} \sqrt{2}\, e_1 & 0 \\ 0 & 0 \end{pmatrix} \,
\end{equation}
illustrating that insufficiency of having a single massless gravitino to exclude a critical point on mWGC grounds.

\section{Spontaneous Supersymmetry Breaking and Goldstino Condensation} \label{sec3}

In the previous section, we considered constraints on (quasi-)de Sitter configurations that arise from the magnetic Weak Gravity Conjecture and the connection between the scalar potential and the gauging in theories with extended supersymmetry. In this section we turn our attention to a more recent result concerning de Sitter constructions that can be formulated in  $\mathcal{N}=1$ supersymmetry models involving nilpotent chiral multiplet. This nilpotent superfield represents the spontaneous supersymmetry breaking sector that appears, for example, on the worldvolume of anti-branes \cite{Bergshoeff:2015jxa,Bandos:2015xnf,Dasgupta:2016prs,Cribiori:2019bfx} and are ubiquitous in de Sitter constructions such as \cite{Kachru:2003aw,Kachru:2003sx,Balasubramanian:2005zx,Bena:2022cwb}.

What we will show in this section is that the dynamics of this supersymmetry breaking sector, as described by the Volkov-Akulov model, include an instability toward the formation of composite states of the goldstino. Depending on the endpoint of this instability, which we currently can not determine, this formation of goldstino composite states can potentially have devastating effects on existing de Sitter constructions that involve anti-branes uplifts. Even in the case of a benign endpoint to the instability, the properties of the resulting configuration are likely to differ from what is commonly assumed and deserves further examination. For the sake of brevity we restrict ourselves to the main ideas and an outline of the calculations involved in deriving this result. All the necessary additional details can be found in \cite{DallAgata:2022abm}.

\subsection{Volkov-Akulov with Lagrange multipliers}

Our starting point is the Volkov-Akulov model \cite{Volkov:1973ix}, which is a theory of a single goldstone fermion with Lagrangian

\begin{equation}
\begin{aligned}
{\cal L} = 
- f^2 
+ i G \sigma^m  \partial_m \overline G 
- \frac{1}{4 f^2} \overline G^2 \partial^2 G^2 
- \frac{1}{16 f^6} G^2 \overline G^2 \partial^2 G^2 \partial^2 \overline G^2
\end{aligned}
\end{equation}
The higher derivative terms are required by supersymmetry, which is realized non-linearly, with the supersymmetry transformation taking the form $\delta_\epsilon G = - \sqrt{2} f \epsilon + ... $.

This model appears, at least as a sub-sector, in the low-energy description of theories with spontaneous supersymmetry breaking. The supersymmetry breaking scale is given by $\sqrt{f}$. The model can alternatively be written in superspace by means of a nilpotent chiral superfield \cite{Cribiori:2017ngp,DallAgata:2016syy}
\begin{equation}
\begin{aligned}
X = \phi + \sqrt{2}  \theta G  +  \theta^2 F \ , \ X^2 = 0  \implies \phi = \frac{G^2}{2F}
\end{aligned}
\end{equation}
with Lagrangian given by
\begin{equation}
\begin{aligned}
L&=\int d^4 \theta K(X, \overline{X}) + \bigg( \int d^2 \theta W(X) + c.c \bigg) \\ 
&= \int d^4 \theta |X|^2 + \left(\int d^2\theta f X + \text{c.c.} \right)
\end{aligned}
\end{equation}
or equivalently, one can impose the nilpotency condition via a second non-dynamical chiral superfield $T = \tau + \sqrt{2}  \theta \lambda  +  \theta^2 B$ that acts as a lagrange multipler
\begin{equation}
\begin{aligned}
L
&= \int d^4 \theta |X|^2 + \left(\int d^2\theta f X + T X^2 + \text{c.c.} \right)
\end{aligned}
\end{equation}
Despite appearing to have additional fields, the scalar fields $\phi, \tau$ as well as the fermion $\lambda$ are all non-dynamical and can be written as an expansion in goldstino multilinears as
\begin{equation}
\begin{aligned}
\langle \phi \rangle \ \sim \ \Big{\langle} \frac{G^2}{f} \Big{\rangle} + \dots  \ , \quad \langle \tau \rangle \ \sim \ \Big{\langle} \frac{\partial^2 \overline G^2  }{f^2} \Big{\rangle} + \dots 
\end{aligned}
\end{equation}
and $\lambda$ can be determined by acting with a supersymmetry transformation on $\tau$. Thus the goldstino is still the unique independent degree of freedom in the theory. 

\subsection{Detecting Composite States with Exact RG}

The presence of goldstino self-interactions presents the possibility that besides the perturbative goldstino states, composite states might also appear in the theory, represented precisely by expressions such as those for $\phi$ and $\tau$ above. One way to determine whether this happens is to allow the theory to evolve along the Wilsonian RG flow and see whether the non-dynamical field $\tau$ acquires a correct-sign kinetic term, similar to Nambu–Jona-Lasinio and composite Higgs models \cite{Nambu:1961tp,Nambu:1961fr,Bardeen:1989ds}. In this case, not only would a kinetic term indicate that $\tau$ itself represents dynamical states, but also the nilpotency condition on $X$ would get relaxed and so $\phi$ would also enter the theory as an independent degree of freedom. The behavior of these new composite states will then be governed by the resulting scalar potential for $\phi$ and $\tau$.

Since the $T$ superfield starts out with a vanishing kinetic term, small variations of the Wilsonian cutoff will produce a small kinetic term, which is equivalent to strong coupling. We can therefore not rely on any perturbative calculation in the couplings and must turn to the formalism of the Exact Renormalization Group \cite{Polchinski:1983gv,Ball:1993zy,Litim:2018pxe}.

The main idea behind this formalism is that given a Wilsonian effective action $L[\Phi ; \mu]$ with cutoff $\mu$, the partition function

\begin{equation}
\begin{aligned}
\mathcal{Z}[\Phi] &= \int \mathcal{D}\Phi \ e^{L_{\text{prop.}}[\Phi ; \mu] + L_{\text{int.}}[\Phi ; \mu] } \\
L_{\text{prop.}}[\Phi ; \mu] &= \int \frac{d^4 k}{(2\pi)^4} \Phi^A(-k) C_{AB}^{-1}(k) \Phi^B(k) \\
L_{\text{int.}}[\Phi ; \mu] &= \int \sum_\lambda g_\lambda(\mu) \prod_{\{ A\}_\lambda} \bigg( \frac{d^4 k_A}{(2\pi)^4} k_A^{n_{A,\lambda}}\Phi^A(k_A) \bigg) \delta(\sum_{ \{ A\}_\lambda } k_A)
\end{aligned}
\end{equation}
should be independent of $\mu$. Here we separated the action into a regularized propagator piece $L_{\text{prop.}}$, which needs not include the full kinetic term, and an interaction piece $L_{\text{int.}}$, which can still have term quadratic in the field. All fields, couplings and momenta are taken here to be rendered dimensionless by rescaling with appropriate powers of $\mu$.

The requirement of invariance of the partition function leads to the condition \cite{Polchinski:1983gv}
\begin{equation} \label{ERG}
\begin{aligned}
\dot{L}_{\text{int.}} &\equiv -\mu \partial_\mu L_{\text{int.} }  \\
&= \int \frac{d^4 k}{(2\pi)^4} \tilde{C}^{AB}(k) \bigg(\frac{\delta^2 L_{\text{int.}} }{\delta \Phi^A(-k) \delta \Phi^B (k) } + \frac{\delta L_{\text{int.}} }{\delta \Phi^A(-k) } \frac{\delta L_{\text{int.}} }{ \delta \Phi^B (k) }   \bigg)  
\end{aligned}
\end{equation}
where $\tilde{C}^{AB}$ is related to $C^{AB}$ in a specified way that depends on the spin of the field. It is worth emphasizing that this flow equation does not make use of any small-coupling approximation. It does however produce an infinite system of equations for all the higher derivative couplings that will generically be generated by the RG flow. Solving these equations thus requires an appropriate choice of truncation of these equations, typically based on a derivative expansion.

For our case, we resort to a truncation of the equations that we dub the Supersymmetric Local Potential Approximation (SLPA). In this approximation we ignore all contributions {\it to and from} any terms that can not be expressed in terms of a K\"ahler or superpotential. In superspace, this amounts to ignoring all terms that are higher-order in superderivatives. The SLPA has the benefit of preserving supersymmetry (assuming a supersymmetric regulator function) and keeping track of the kinetic terms of the scalar fields, which is what we are interested in. This is in contrast to the more common Local Potential Approximation (LPA) which would simply truncate all derivative interactions \cite{Zumbach:1994vg,Zumbach:1994kc,Morris:1994ie,Harvey-Fros:1999qpe}.

The truncation of higher order terms means that we should not trust the solutions to the remaining equations for large changes in $\mu$. This can be an issue when the qualitative features one is after do not appear immediately in the RG flow, such as the tachyonic behavior in composite Higgs models \cite{Bardeen:1989ds}. As we will see, the features  describing here will not suffer from this problem, and will be immediately visible even for a small decrease in $\mu$.

In order to apply the ERG formalism to the Volkov--Akulov model, we start with the following K\"ahler and superpotentials

\begin{equation} \label{kahsupot}
\begin{aligned}
K &= \alpha |X|^2 + \beta |T|^2 + g \, |T|^2 |X|^2 + \frac14 q \, |X|^4  \\
W &= f X + \frac12 T X^2
\end{aligned}
\end{equation}
and separate $K$ into a (regularized) propagator and interaction part
\begin{equation}
\begin{aligned}
K_{\text{prop.}} &= c^{-1} |X|^2 + c^{-1} |T|^2   \\
K_{\text{int.}} &= (\alpha - 1) |X|^2 + (\beta - 1) |T|^2 + \gamma \mu^{-2} |T|^2 |X|^2 + \frac14 \zeta \mu^{-2} |X|^4
\end{aligned}
\end{equation}
where we assume the regulator function can be expanded as $c(p^2/\mu^2) = 1+ \sum c_n \  p^{2n}/\mu^{2n}$. This action contains all the terms that will be generated by the RG flow within the SLPA. Since the action is manifestly supersymmetric, we can read off the flow of the couplings $\alpha, \beta, \gamma$ and $\zeta$ from any of the terms that contain them. A convenient choice are the quadratic terms in the auxiliary fields $F$ and $B$. The calculation is laborious but straightforward and can be found in full detail in \cite{DallAgata:2022abm}. Here we simply quote the resulting flow equations
\begin{equation}
\begin{aligned}
\dot \zeta &= - 2 \zeta - 2 c_1 \,, \quad 
\dot \gamma = - 2 \gamma - 2 c_1 \,  \\
\dot \beta  &= - 2 N \gamma \, , \quad
\dot \alpha  = - 2 N (\gamma + \zeta) \ ,
\end{aligned}
\end{equation}

The flow of the 4-point couplings $\zeta$ and $\gamma$ is generated by the second term in \eqref{ERG} from the $TX^2$ term in the superpotential. Once these terms are generated, the first term in \eqref{ERG} generates corrections to the kinetic terms from these 4-point couplings.

\subsection{Goldstino condensation in Volkov-Akulov and Beyond}

\begin{figure}
\includegraphics[scale=0.45]{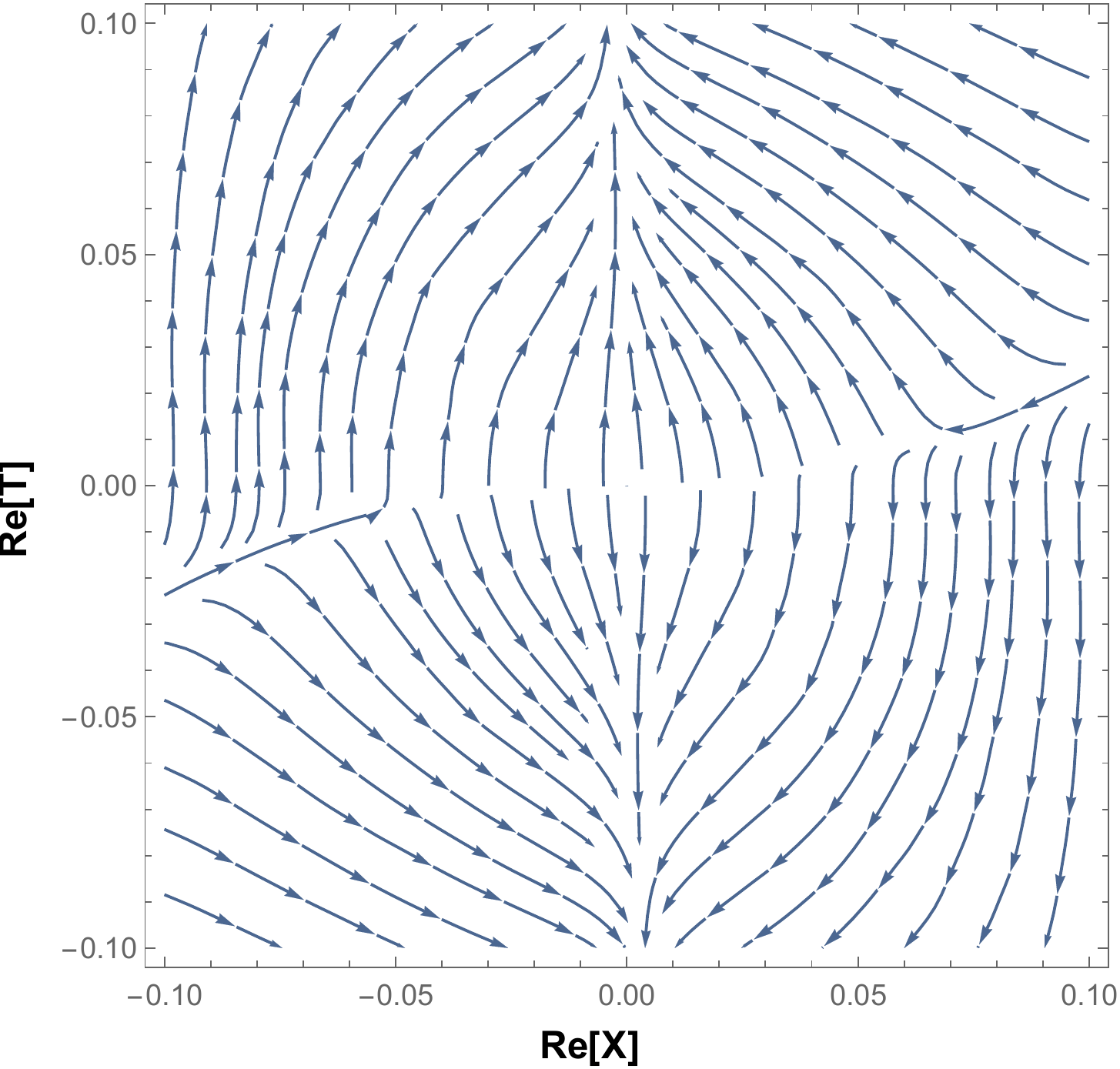} ~~~~ \includegraphics[scale=0.48]{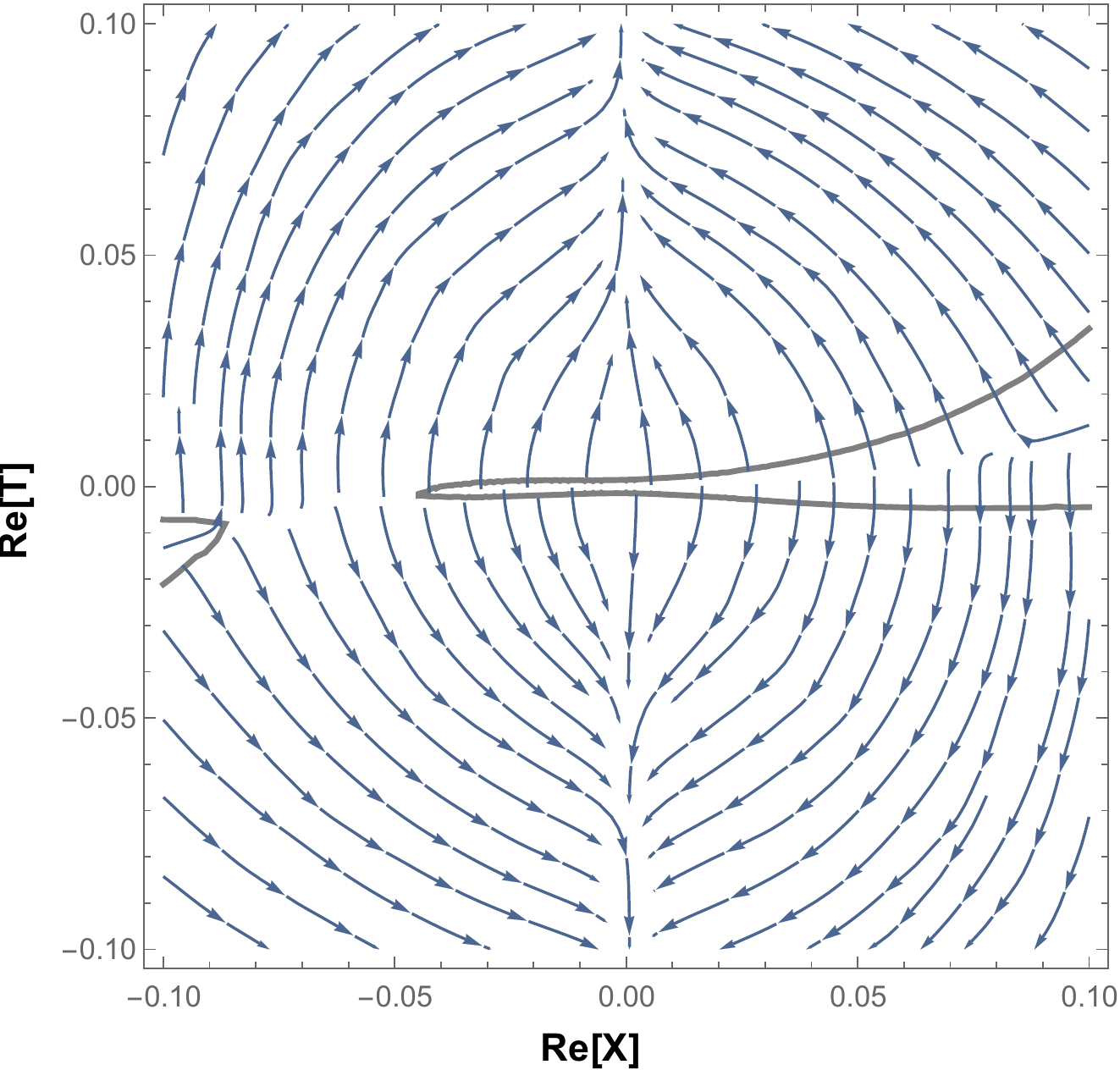}
\centering
\caption{Stream-plot of the (negative) gradient of the scalar potential for pure Volkov--Akulov (left) and for unwarped coupling to KKLT (right) at $t = 0.1$ restricted to the real parts of $X$ and $T$ scalar components. The regulator is chosen to be $c=(1-p^2)\Theta(1-p^2)$ and the UV matching scale is $\Lambda = \sqrt{f}$. The KKLT model parameters are those of the example in \cite{Kachru:2003aw}. The black contour denotes the locus of vanishing gradient along the KKLT K\"ahler modulus whose value is taken to be that of the critical point.}
\label{fig2} 
\end{figure}

We now need to solve the Volkov--Akulov ERG equations with boundary conditions
\begin{equation}
\alpha\Big{|}_{t = 0} = 1 \,, \quad 
\beta\Big{|}_{t = 0} = 0 \,, \quad 
\gamma\Big{|}_{t = 0} = 0 \,, \quad 
\zeta\Big{|}_{t = 0} = 0 \,   
\end{equation}
with $t = \log(\Lambda/\mu)$, where $\Lambda \leq \sqrt{f}$ is a matching UV scale where we might imagine all other degrees of freedom are integrated out. $N$ is the result of a momentum loop integral that appears in the evaluation of the first term in \eqref{ERG} and $c_1$ is the first coefficient in the expansion of the regulator function. A typical monotonic regulator will have $c_1<0$ and $N<0$.
The solution to this system of ODE's is
\begin{equation} \label{ERGsoln}
\begin{aligned}
\zeta &= - c_1 \left( 1 - e^{-2t} \right)   \,, \quad 
\gamma = - c_1 \left( 1 - e^{-2t} \right) \, ,\\
\alpha &= 1 - 2 c_1 N + 4 c_1 N \left(t + \frac12 e^{-2 t} \right) \, , \\
\beta &= - c_1 N + 2 c_1 N \left(t + \frac12 e^{-2 t} \right) > 0 \ . \,  
\end{aligned}
\end{equation}
The $T$ field, which started its life as a Lagrange multiplier, acquires a positive kinetic term, indicating that the composite states of the goldstino that it represents enter the effective theory as independent degrees of freedom. Furthermore, we can evaluate the mass spectrum around the original Volkov--Akulov point at $X=T=0$ and we find that the masses take the form
\begin{equation}
\begin{aligned}
    m_{\pm}^2 = - \tilde{f}^2\left[ \left(\tilde{\gamma} + 4 \tilde{\zeta}\right) \pm \sqrt{\frac{16 \ \tilde{g}^2}{\tilde{f}^2} + \left(\tilde{\gamma} - 4 \tilde{\zeta}\right)^2} \ \right] \,.
\end{aligned}    
\end{equation}
where $\tilde{f}, \tilde{\gamma}, \tilde{\zeta}, \tilde{g}$ are rescaled couplings obtained after canonically normalizing the kinetic terms and are all positive. It isn't hard to see that at least one of the masses is always negative and thus at least one of the scalars will develop an expectation value, representing the formation of a goldstino condensate. The typical behavior of the scalar potential for small $t$ is depicted in the left side of Figure \ref{fig2}.

For larger $t$ contributions beyond the SLPA will alter the flow from that given by \eqref{ERGsoln}. In particular additional couplings will enter the K\"ahler potential and affect the shape of the potential away from the central point. This may ultimately affect the endpoint of the instability and its physical interpretation. The central instability, however, is expected to survive, since both the generation of the kinetic term for $T$ and the tachyonic behavior at the origin is due to the $TX^2$ term in the superpotential. To remove the central instability, the effect of the higher-derivative contributions in the ERG flow would have to give corrections to the K\"ahler potential that compete with the effect of the superpotential term, which would indicate a breakdown of the EFT description altogether. While this is not ruled out, it would require an even bigger revision of what is commonly assumed about the IR behavior of the Volkov--Akulov model. In either case, an analysis beyond the SLPA is called for.

So far we have only considered the pure Volkov--Akulov model without additional fields and in the rigid limit. Additional fields are unlikely to change the qualitative features found above. This is, again, due to the fact that the $TX^2$ term in the superpotential is responsible for both the kinetic term of $T$ and for the presence of the central tachyon. At the UV matching scale, $T$ only appears in that single superpotential term and can not couple to any other fields in any larger model. The SLPA calculation for the kinetic term will therefor be unaffected. Beyond the SLPA, additional fields may have an effect and considerations of the previous paragraph apply. Once again, analysis to higher orders remains a necessary future direction.

As far as extending this analysis to supergravity, a major obstacle is the absence of a supersymmetric regulator, which could be used to maintain a manifestly supersymmetric RG flow. This remains an open problem. That said, the natural expectation is that corrections to the rigid limit results will be suppressed by powers of $M_p$ and so the main qualitative features of a dynamic $T$ and a tachyonic critical point should remain. As a naive first check, one can simply embed the RG-evolved V--A model into the KKLT scenario by adding the RG-evolved K\"ahler and superpotentials \eqref{kahsupot} evaluated at small $t$ to the pre-uplift KKLT K\"ahler and superpotentials. The details of this calculation are also described in \cite{DallAgata:2022abm} and the result is depicted in the right side of Figure \ref{fig2}. The main result is that the dS critical point shifts in the $X,T$- plane and also develops a tachyonic instability toward goldstino condensation. Of course one can look at more complicated models where the goldstino sector couples to other matter fields in more sophisticated ways, such as \cite{Kachru:2003sx,Bena:2018fqc,Dudas:2019pls,Bento:2021nbb}, but the instability of the original critical point toward goldstino condensation is expected to remain for the reasons outlined above.

Finally, we note that since the Volkov--Akulov model is generally understood to be contained in the low-energy description of the worldvolume dynamics of anti-branes, the instability we find should be present in the simple case of the $\overline{D3}/O3$ system in 10 dimensions. In this context, a string theoretic analysis of the goldstino composite states and their dynamics could give insights into the eventual fate of the instability that would be difficult to access using effective field theory methods.

\section{Conclusion}

In this contribution, we have explored potential obstacles to constructing de Sitter spacetimes in supersymmetric effective theories as explored in \cite{DallAgata:2021nnr,DallAgata:2022abm}. We have shown that in $\mathcal{N}=2$ supergavity, (quasi-)de Sitter configurations with massless charged gravitini necessarily have a Hubble scale on the order of, or above, the UV cutoff dictated by the magnetic Weak Gravity Conjecture, thus invalidating their EFT description. This constraint excludes many of the known de Sitter critical points in $\mathcal{N}=2$ supergravity, including all known stable ones. Possible future directions in this area include clarifying the limits of applicability of the mWGC criterion to time-dependent backgrounds, in particular those that pass near critical points with non-abelian enhancement of the gauge group. Investigations of constraints coming from the charges of other particles, particularly at points where the gravitini are uncharged. These two lines of analysis could help decide the fate of the critical points encountered in the second example of section \ref{sec2}.

We have then turned our attention to models where supersymmetry breaking is achieved through an explicit goldstino sector where supersymmetry is non-linearly realized. We have shown, using exact renormalization group techniques, that the pure Volkov--Akulov model has an instability toward goldstino condensation and argued that additional matter or supergravity couplings are unlikely to remove the instability, but may affect its endpoint. In either case, this new effect invites us to re-examine the properties of string models involving anti-branes. Future directions in this area include moving beyond the SLPA to determine the possible endpoint of the instability as well as analyzing the composite states in the context of the $\overline{D3}/O3$ system using worldsheet or string field theory methods.

\section*{Acknowledgements}
\noindent This work is supported by the STARS grant SUGRA-MAX. 

% BIBIB

\end{document}